\documentclass[12pt]{article}
\def\be{\begin{equation}} \def\ee{\end{equation}}
\def\bea{\begin{eqnarray}} \def\eea{\end{eqnarray}} \def\ba{\begin{array}}
\def\ea{\end{array}} \def\ben{\begin{enumerate}} \def\een{\end{enumerate}}

\newcommand{\eqn}[1]{(\ref{#1})}

\newcommand{\hepth}[1]{{\tt hep-th/{#1}}}

\def\l{\lambda}
\def\m{\mu}

\def\n{\nu}

\def\br{\nonumber\\}
\def\ud{\underline}

\begin{document}
{}~
\hfill\vbox{\hbox{hep-th/yymm.nnnn} \hbox{\today}}\break

\vskip 3.5cm
\centerline{\large \bf  D2-D8 system with  massive strings}
\centerline{\large \bf  and }
\centerline{\large \bf   the Lifshitz spacetimes}

\vskip 1cm

\vspace*{.5cm}

\centerline{  Harvendra Singh}

\vspace*{.25cm}
\centerline{ \it  Theory Division, Saha Institute of Nuclear Physics} 
\centerline{ \it  1/AF, Bidhannagar, Kolkata 700064, India}
\vspace{.2cm}
\centerline{ \it  Homi Bhabha National Institute, Anushakti Nagar, Mumbai 400094, India}
\vspace*{.25cm}

\vspace*{.5cm}

\vskip.5cm

\vskip1cm

\centerline{\bf Abstract} \bigskip
The  Romans' type IIA supergravity
 allows  fundamental strings to have explicit mass term 
at the tree level. We show that there exists a  (F1,D2,D8) 
 brane configuration which  gives rise to  
$Lif_4^{(2)}\times {R}^1\times S^5$ vacua 
supported by the massive strings. 
The presence of D8-branes  naturally excites  massive fundamental strings. 
A compactification  on  circle relates  these  Lifshitz massive type-IIA
background with the axion-flux $Lif_4^{(2)}\times {S}^1\times S^5$ 
vacua in ordinary type-IIB theory. 
 The massive  T-duality  in eight dimensions further relates 
them to yet another  $\widetilde{Lif}_4^{(2)}\times S^1\times S^5$  vacua 
constituted by  (F1,D0,D6)  system 
in ordinary type IIA theory. The latter vacua after compactification 
to four dimensions  generate  
two `distinct' electric charges
and a constant magnetic field, all living over 2-dimensional plane. 
This somewhat
reminds us of a similar set up in  quantum Hall systems.

\vfill 
\eject

\baselineskip=16.2pt

\section{Introduction}
The AdS/CFT  \cite{Maldacena:1997re,Gubser:1998bc, Witten:1998qj}
 holgraphic applications in strongly 
coupled quantum systems exhibiting nonrelativitic symmetries near critical points
 has been the focus of studies recently \cite{son}-\cite{taylor}.  
Some of the holographic applications may involve   strongly coupled fermionic 
systems at finite density or a gas of ultra cold atoms 
\cite{son, bala}. Some  related framework of Galilean symmetry were 
studied even earlier  \cite{horvathy}.  For 
finite temperature properties such as phase transitions, 
transport and viscosity we include black-holes in asymptotically 
AdS  backgrounds. For superconductivity phenomenon  the 4-dimensional
non-relativistic  geometry generically involves 
spontaneously broken Higgs phase where the  Maxwell field 
is massive \cite{herzogrev,denefrev}. We shall discuss here some 
examples of 10-dimensional Lifshitz spacetimes
 where a Higgs phase instead involves  2-rank 
antisymmetric tensor field and has massive fundamental strings. 
The two phenomena 
indeed have parallel from 10-dimensional perspective. It is because
a Kaluza-Klein compactification of a (massive) 2-rank tensor field
on a circle  gives rise to a (massive) gauge field in  lower dimensions.

Our main interest in this work is to construct Lifshitz  solutions with dynamical 
exponent $a=2$, directly in  Roman's  type IIA supergravity  \cite{roma}.
The massive type IIA theory is the only known example of a 
10-dimensional maximal supergravity where the string 
field is explicitly massive at the tree level. 
Thus the massive type IIA  provides an unique setup to look for Lifshitz  
and Schr\"odinger  like
 solutions which involve massive fields and study  their dual
nonrelativistic field theories on the boundary.
There are no prior
attempts to our knowledge where the same has been worked out for 
Romans type IIA supergravity, however $Sch_4^{(3)}\times S^6$ massive string
vacua are known to us \cite{Singh:2009tq}. 
The $Lif_4^{(2)}\times R^1\times S^5$ have been shown to exist 
in $10d$ type IIB
string theory when appropriate axion flux is switched 
on \cite{Balasubramanian:2010uk}. We provide an exlicit example of 
$Lif_4^{(2)}\times R^1\times  S^5$ background,
which is  a solution of massive type IIA and it does involve massive 
$B$-field. It is a background 
generated by a bound state of $(F1,D2,D8)$ brane system. 
It is important to know all such vacua
as their compactification  on $S^1\times S^5$ 
immediately provides the prototype  
$Lif_4^{(2)}$ background in $4d$,  which is proposed to 
be dual to non-relativistic Lifshitz  theory with dynamical exponent two. 
By exploiting massive T-duality symmetries in $8d$ we further 
provide another example of 
$\widetilde{Lif}_4^{(2)}\times S^1\times  S^5$ background, 
which is generated by massless
$(F1,D0,D6)$ brane system in ordinary type IIA theory. 
     
The paper is organised as follows. In the section-2 we first  review the 
relevant aspects of massive type IIA  sugra action and its known Schr\"odinger
solution. In section-3 we write down new  Lifshitz
solution generated by the $(F1,D2,D8)$ system and 
discuss  subtle aspects under massive T-duality for these solutions. 
The section-4 contains  ordinary type IIA solution involving
$(F,D0,D6)$ branes,
which is supported by RR 2-form flux and massless $B$-field. They 
have a very crucial $(B_{(2)}\wedge G_{(2)})^2$  coupling between them.
The string $B$-field gets essentially nested with constant magnetic 
flux of $G_{(2)}$ over $2d$ plane. 
 We next discuss their consistent  reduction to four and five dimensions 
and their effective bulk
theories in section-5. We provide an M-theory uplift of these 
vacua in section-6. A summary is provided in the section-7.

\section{$Sch_4^{(3)}\times S^6$: massive strings }

In this section we mainly review some useful information about 
 10-dimensional massive type IIA supergravity  \cite{roma} and a known
Schr\"odinger vacua $Sch_4^{(3)}\times S^6$  \cite{Singh:2009tq}. 
One may choose to directly skip to the next section. 
 The Romans type IIA theory is the only known maximal 
supergravity in ten dimensions which allows massive string $B$  field.
The theory is described  by the following bosonic action
\bea
S={1\over G_N}\int  \bigg[e^{-2\phi}\left(  \ast R + 4
(d\phi)^2 
-{1\over 2} (H_{(3)})^2  \right)
-{1\over 2} (G_{(2)})^2  
-{1\over 2} (G_{(4)})^2  
-\ast{m^2\over 2}  \bigg]
\eea
where  topological terms have been dropped because
these would be  vanishing 
for the Lifshitz backgrounds we shall be studying in this paper, see 
for details in \cite{roma, hs2001}.\footnote{
We are adopting a convention: $
\int (H_{(p)})^2=\int H_{(p)}\wedge\ast H_{(p)}
={1\over p!}\int d^{10}x \sqrt{-g}
H_{\mu_1\cdots\mu_p}H^{\mu_1\cdots\mu_p}$ and for  scalar quantities 
like curvature scalar: 
$\int \ast R=\int d^{10}x \sqrt{-g} R$  .}  The field strengths 
are defined as
\bea
H_{(3)}=dB_{(2)},~~~G_{(2)}=dC_{(1)} +m B_{(2)}, ~~~
G_{(4)}=dC_{(3)} +B_{(2)}\wedge dC_{(1)} + {m\over2} 
B_{(2)}\wedge B_{(2)}
\eea
where $m$ is the mass parameter and positive cosmological constant of
the 10-dimensional theory. The cosmological constant however
 generates   a nontrivial  potential term  for the dilaton 
field. Other than the nonsupersymmetric 
Freund-Rubin  vacua in \cite{roma}, some 
 known BPS solutions of the  theory 
include D8-branes \cite{Pol1, berg, wittd0d8,park00,ohta01,hull}, the
$K3$ compactifications \cite{haack}, the $(D6,D8)$  and  $(D4,D6,D8)$ 
bound states \cite{hs2001, hs1}.
Under the `massive' T-duality \cite{berg} the D8 brane  
 can be mapped over to a D7 brane in ordinary type IIB string theory.  
The string $B$-field  is explicitly massive with mass square given by $m^2$,  
and it plays an important role in obtaining
nonrelativistic Schr\"odinger solutions \cite{Singh:2009tq}. 
The massive type IIA theory however  never admits a flat Minkowski vacuum 
solution. But in the $m\to 0$ limit the massive
theory reduces to ordinary type IIA supergravity. 
   
An observed common  feature in four-dimensional AdS gravity theories
has been that in order to obtain 
Schr\"odinger or Lifshitz type non-relativistic 
solutions one needs to include 
massive (Proca) gauge fields in the action \cite{son,bala}. 
(Although massless gauge fields  can give rise to  nonrelativistic vacua 
however,  in  simple cases of  D-branes
compactified along lightcone coordinate \cite{hs2010,hs2012}, 
they usually give rise to  conformal (or hyperscaling) 
Lifshitz or Schr\"odinger vacua.)   
Particularly for massive
type IIA theory  the existence of  Schr\"odinger solution
 $Sch_4^{(3)}\times S^6$ has already been shown in  
\cite{Singh:2009tq}. 
In the rest of this section we  review the $a=3$ Schr\"odinger solution 
to familiarise ourselves, and also because these vacua 
are  constituted by  D0, D2, D8 branes, alongwith massive strings
which we will encounter  again when we write down Lifshitz solutions
of the theory.

The massive type IIA action in the tensorial notation is given by
\bea\label{eq2l}
S={1\over G_N}\int d^{10}x \sqrt{-g}\bigg[ R- {1\over2} 
(\partial_\m\phi)^2 
-{e^{-\phi}\over 2.3!} (H_{\m\n\l})^2  
-{e^{3\phi/2}\over 2.2!} (G_{\m\n})^2 \br  
-{e^{\phi/2}\over 2.4!} (G_{\m\n\l\sigma })^2  
-{e^{5\phi/2}\over 2} m^2 \bigg],
\eea
with the field strengths $H_{\m\n\l}=3\partial_{[\m} B_{\n\l ]},~ 
G_{\m\n}=2\partial_{[\m}C_{\n]} + m B_{\m\n} ,~
G_{\m\n\l\sigma}=4\partial_{[\m} C_{\n\l\sigma ]}+ 6
B_{[\m\n}G_{\n\sigma ]} -
3{m}B_{[\m\n}B_{\n\sigma ]}$.
The equations of motion of this theory admit following
  $Sch_4^{(3)}\times S^6$ solution \cite{Singh:2009tq}
\bea\label{sol2}
&&ds^2=L^2\left(-{2\over z^{6}} 
(dx^{+})^2+{-2dx^{+}dx^{-}+dy^2+dz^2\over
z^2}  +{5\over 2} d\Omega_6^2 \right) ,\br
&&e^\phi=g_a, ~~~C_{+-y}= -\sqrt{10}  g_a^{-{1\over 4}} {L^3\over 3 z^3}, 
~~~C_{+}= -{\sqrt{5}\over 3}{ g_a^{-3/4}L\over  z^3} \br
&& 
B_{+ y}=  -{g_a L^2\over 2z^4} \ ,
\eea 
 where $L^2={2\over  m^2 g_a^{5/2}}$. But we should have string coupling
$g_a \ll 1$ and $L\gg 1$.
The radius of curvature of the sphere is directly related to the mass parameter.
Note that the Lorentz invariance is explicitly broken in the solutions, although 
spin group of the sphere is  intact. The Schr\"odinger vacua \eqn{sol2}  
involves a collection of D0, D2, and 
D8 branes (which are wrapped around $S^6$), along with massive 
fundamental strings (F1) stretched along  $y$ direction.  
The Einstein equations do
 involve a nontrivial stress-energy tensor component
$T_{++}$ that  receives contributions from the 
lightlike  components of these fields. 
From the solution we learn that in ten-dimensional sense
the  matter  (dust) responsible for a Sch\"odinger 
solutions is  made up of D-branes and most importantly
the {\it massive}  strings. 
The boundary of the spacetime is located at $z=0$, and near to
the boundary 
the metric and fields become divergent, which usually is the case for
non-relativistic vacua with $a>1$. But everything is fine in 
the interior of the spacetime including the curvature scalar 
which is a constant quantity. 

\section{$Lif_4^{(2)} \times R^1 \times S^5$ vacua}
\subsection{D2-D8: tied up with massive strings}

It is  known that the
 Freund-Rubin  $AdS_4\times S^6$ maximally symmetric
vacua in Romans theory \cite{roma} 
is constituted by $(D2,D8)$ branes. But there are no `massive'  
$B$-field excitations in them.    
We  show  that one can  construct  Lifshitz   vacua 
that are  supported by  massive $B$-field, D2 and D8 branes.
The strings in these solutions becomes massive after gobbling up all 
D0 branes, unlike the Schr\"odinger type vacua \eqn{sol2} above 
where some D0 branes are still present. 
Thus the D0 brane charges  will not be  explicitly seen 
in the following Lifshitz solutions.
These higgsed Lifshitz solutions are given by (in string  metric) 
\bea\label{sol2a}
&&ds^2=\alpha' L^2\left(- {dt^2\over  z^4} +{dx_1^2+dx_2^2\over z^2}+{dz^2\over
z^2}  +{dy^2\over  q^2} + d\Omega_5^2 \right) ,\br
&&e^\phi=g_a, ~~~~~C_{(3)}= -{(\alpha' L^2)^{3\over 2} \over g_a}
{1\over  z^4}dt\wedge dx_1\wedge dx_2, \br &&
B_{(2)}=  {\alpha' L^2\over q z^2}dt\wedge dy  \ ,
\eea 
with  $L={2\over g_a  m l_s}$, where $m$ being the mass 
parameter in the Romans' action. The $q$ 
is  free (length)  parameter 
and $g_a$ is  perturbative string coupling in this  massive type IIA vacuum. 
Note $L$, which is dimensionless,
determines overall radius of curvature of the 10-dimensional spacetime.
While $m$ being a parameter in the lagrangian and it determines 
 $L$,  therefore Romans' theory with 
  $m \ll{ 2\over g_a l_s}$ would be preferred 
so that we can have $L\gg 1$ in  the solutions \eqn{sol2a}, else
  we cannot trust these classical vacua. \footnote{Note, from  the D8
brane/domain wall correspondence \cite{berg}, one typically
expects $m \approx {g_a N_{D8} \over l_s}$, a value which is definitely
well within ${ 2\over g_a l_s}$.}

The Lifshitz configuration \eqn{sol2a} describes a parallel stack 
of D2-branes stretched along $(x_1,~x_2)$
directions and `massive' fundamental strings that are aligned
along  $y$ direction. The D8-branes  wrap around $S^5$ 
completely while remaining stretched along the  patch $(x_1,~x_2,~y)$. 
The  D2 charges can be found as
\be
\label{d2q}
Q_{D2}\sim{1\over\omega_5}
\int e^{\phi\over 2} \star G_{(4)}
\sim {4l_s^5 L^5 l_y\over g_a q}  \propto  N 
\ee
where $l_y$ is coordinate size of $y$  and $\omega_5$ is the unit
 volume of  5-sphere. 
Thus eq.\eqn{sol2a} describes a  $(F1,D2,D8)$ configuration, in which
D2 branes are  stacked inside D8 worldvolume. The D2 stack is studded (threaded)
by massive F-strings, having mass  $m={2\over L g_a  l_s}$. The D0 branes 
are all gauged away (or eaten up) by the F-strings which  have become  massive.
This phenomenon  happens due to  the higgsing  (or stueckelberg mechanism) 
 in massive type IIA theory \cite{roma}.    
In these solutions $y$ remains an overall isometry direction which 
may be  compact also. The holographic
$z$ coordinate  acts as a common
transverse direction for all the branes. 

\subsection{ $Lif_4^{(2)} \times S^1_w \times S^5$}
The 10-dimensional string coupling  $g_a$ 
has to be weak so that we can trust the Lifshitz   vacua  \eqn{sol2a}. 
The small $g_a$ however   tends to send the mass of  F-strings
to higher values, but for sufficiently  large  
$L\sim O({1\over g_a^2})$ it
can  support  massive strings in the background (note the lighter ones with 
$m\approx {g_a\over  l_s}$ only need to be excited). 
So far we considered $y$ as  noncompact coordinate
and $q$  an associated arbitary 
length scale. Let us consider the case 
when  $y$ is a compact circle, $i.e.$ $y\sim y+2\pi R_y$. 
The physical radius of  $y$-circle will be $ {l_s L R_y \over q}$. 
Therefore for smaller $q$  values the massive strings can  indeed be excited 
in the transverse  $y$ direction of the D2-branes. It would  also
be appropriate here to take 
$q\equiv w R_y$,  with $w$ being an integer. (It could also be
the wrapping number of D8-branes. By wrapping number here we mean by 
number of times a single D8 brane wraps around $S^1$.) 
The vacua \eqn{sol2a}  will now be described as 
$Lif_4^{(2)} \times S^1_w \times S^5$:
\bea\label{sol2ae}
&&ds^2=\alpha' L^2\left(- {dt^2\over  z^4} +{dx_1^2+dx_2^2\over z^2}+{dz^2\over
z^2}  +{1\over  w^2} d\psi^2 + d\Omega_5^2 \right) ,\br
&&e^\phi=g_a, ~~~C_{(3)}= -{(\alpha' L^2)^{3\over 2} \over g_a  z^4}dt\wedge dx_1\wedge dx_2, ~~~~
B_{(2)}=  {\alpha' L^2\over w z^2}dt\wedge d\psi  \ ,
\eea 
where $0\le \psi\le 2\pi$. 
($\psi$  can also be viewed as an orbifolded circle.) 
 Anyhow $w>1$ tells us about the comparative sizes of $S^1$ and $S^5$. 
Due to this size difference a D8-brane can wrap $S^1$ $w$-times more 
as compared to its wrapping around $S^5$. 
Especially for $\omega=1$, 
both $S^1$ and $S^5$ will have the same size. 
But small $w$  is always preferred in \eqn{sol2ae}. 
For large $w$ (or $q$)  the radius of $S^1$ will  
 become sub-stringy, it would then be appropriate to switch over to the 
T-dual  vacua in type IIB string theory. 
We would discuss it in the next section.       

The  Lifshitz solutions  \eqn{sol2a} or \eqn{sol2ae} both have  
following asymmetric scaling  symmetry
\be
t\to \lambda^2 t,~~
x^{i}\to \lambda x^{i},~~
z\to \lambda z
\ee
Note $y$ (or $\psi$)  coordinate is not required to scale at all 
and it is one of the charactersitics
of the $a=2$ Lifshitz vacua.

\subsection{Massive $9d$ T-duality }

Upon $S^1$ compactification   the Romans 
 theory gives rise to a massive supergravity  in  nine dimensions. 
Keeping only the  fields relevant for our background \eqn{sol2a}, 
(a complete circle compactification can be found  \cite{berg})
a $9d$ massive action can be written as 
\bea\label{eq2}
S\sim\int d^{9}x \sqrt{-g}e^{-2\phi_9}\bigg[ R
+ {4} (\partial_\m\phi_9)^2 
- {1\over2} (\partial_\m\rho)^2 
-{e^{\sqrt{2}\rho}\over 2.2!} (F_{\m\n})^2  
-{e^{\rho\over \sqrt{2}}\over 2} m^2 (A_{\m})^2 \br  
-{e^{-\rho\over\sqrt{2}}\over 2.4!} (G_{\m\n\l\sigma })^2  
-{e^{-\rho\over\sqrt{2}}\over 2} m^2 \bigg],
\eea
where the vector field $A_\mu$ which arises directly
from reduction of massive tensor component 
$\hat B_{\mu\ud{y}}$ is only kept in the action.
The $9d$ dilaton field is given by
$$\phi_9=\hat\phi-{1\over 2\sqrt{2}}\rho $$ 
whereas
 $$~\rho= {1\over \sqrt{2}}\log (\hat{g}_{yy})$$
 defines the radion mode along $y$. 
(The $10d$ fields are denoted with a hat sign to distinguish them 
from 9-dimensional ones.) The 
Kaluza-Klein gauge field  is set to vanish. 
The topological terms in the action are also ignored as these
are not  relevant for the Lifshitz background  
we are studying here. By setting $\alpha'=1$, 
 after compactification along $y$ the eq.\eqn{sol2a}  reduces to 
\bea\label{sol2b}
&&ds^2_{9d}=L^2\left(- {dt^2\over  z^4} +{dx_1^2+dx_2^2+dz^2\over
z^2} +  d\Omega_5^2 \right) ,\br
&&e^{\phi_{9}}=\sqrt{q\over L} g_a  ,  
~~~C_{(3)}= -{L^3\over g_a z^4}dt\wedge dx_1\wedge dx_2 \br
&& 
A_{(1)}= {L^2\over q z^2}dt, ~~~ e^{-\sqrt{2}\rho}=q^2 L^{-2}\ ,
\eea 
It can be checked that \eqn{sol2b}
 is a consistent solution of  the $9d$ action   \eqn{eq2}. 
Actually both the dilaton $\phi_9$ and the $\rho$ have 
constant background values.
Since it is a solution of $9d$ type II sugra with a massive vector field, 
the above $9d$ Lifshitz vacua
 can  be uplifted back  to 10-dimensional  type IIB theory
(over a dual radius circle) by exploiting `massive' T-duality \cite{berg}.
By employing the massive duality rules \cite{berg,michaud}  we 
get following type IIB solution
\bea\label{sol2c}
&&d{s}^2_{IIb}=L^2\left( {\tilde{q}^2}{d\tilde{y}^2} - 
{2
\tilde{q}dt d\tilde{y}\over z^2} +{dx_1^2+dx_2^2\over z^2}+{dz^2\over
z^2}   + d\Omega_5^2 \right) \br
&&
G_{(5)}= {1\over \sqrt{2}g_b}(
{4\tilde{q}L^4\over  z^5}dt\wedge dx_1\wedge dx_2\wedge d\tilde{y}
\wedge dz + {\rm self~dual}) \br
&&e^{\phi_b}=g_b , ~~~~ \chi= {2\tilde{q}\over g_b} \tilde y \ .
\br 
\eea 
The type IIB string coupling $g_b$, radius of curvature $L$, 
and flux $\tilde{q}$ are all independent parameters.
The  coupling $g_b$  should be kept small.
The parameter $\tilde{q}$  determines the required
axion flux  and also the relative size of $\tilde {y}$ in the metric. 
(Recall  an ordinary type IIB  theory has 
no mass scale  of its own unlike the Romans' type IIA theory.)
The massive $9d$ T-duality  does indeed
relate $\tilde{q}$ with Romans' mass through
\be 
g_b=g_a {q\over L}, ~~~m={2\over g_a L}\equiv{2\tilde{q}\over g_b},~~~
\tilde{q}\equiv {q\over L^2}
 \ee
Note the  flux parameter $\tilde{q}$ has  mass dimension one.
The vacua \eqn{sol2c} describes   Lifshitz solution
 of  type IIB  theory first obtained by 
\cite{Balasubramanian:2010uk},
if we redefine the coordinates as $\tilde{y}=x^+, t=x^-$.

Note  that  the axion field has a  linear dependence 
on the circle coordinate ($\tilde y\sim \tilde y +2\pi \tilde{R}_y$), 
where $\tilde{R}_y={\alpha' L^2\over R_y}$
via T-duality. Due to this the axion 
undergoes discrete jumps (determined by parameter $\tilde{q}$) each time  
it goes around the circle, as it has to be periodic.
One should  take $\tilde{q}={w R_y\over L^2 }$ with 
an integral $w$. The 
$w$ will now effectively count the  axions  (or the number of D7 branes)
\be 
Q_{D7}={1\over 2\pi}\int d\chi ={\tilde q}{\tilde R_y}=w\ .
\ee
The  D3 charge is given by
\be
\label{d3q}
Q_{D3}\sim{1\over \omega_5}
\int  \star G_{(5)}
= {2\sqrt{2} L^4 \over g_b}  \approx  N 
\ee
taking  $L^4\approx g_b N$. Thus the eq.\eqn{sol2c} 
 describes a bound state of  $(W,D3,D7)$  system with wave.
The  coordinates $z$ and $\tilde y$ together constitute 
two transverse directions of D7 branes.   
It is  the same kind of effect  which the massive T-duality 
brings under which D8-brane of massive type IIA is 
mapped over to D7-brane in type IIB and vice-versa \cite{berg}.
Thus we have 
confirmed that the Romans type IIA theory with  massive  $B$-field
admits   $(F1,D2,D8)$ Lifshitz vacua and it also 
provides a consistent T-dual  description of the  
 $(W,D3,D7)$ Lifshitz vacua of type IIB string theory.

The vacua presented here are different from 
other Lifshitz vacua obtained in $6d$ and $5d$ 
gauged/massive supergravities with arbitrary dynamical exponents
in \cite{greg,greg1}. Those 
$Lif_4^{(a)}\times H_2$ and $Lif_3^{(a)}\times H_2$ vacua 
with $a>1$ (the $H_2$ being sphere, hyperboloid or  flat space) 
can also be lifted to $10d$, but these are different cases.
Only dilatonic scalars are present in those 
gauged/massive sugra theories. We are note sure if they could
 be ralated to $(F1,D2,D8)$ background.

\subsection{$5d$ massive tensor model}

After doing an explicit $S^5$ 
 compactification the 10-dimensional background  \eqn{sol2a}
reduces to the following 5-dimensional 
$Lif_4^{(2)}\times R^1$ solution (we set $\alpha'=1$ for simplicity)
\bea\label{sol2vn3gg}
&&ds^2={L}^2\left(- {dt^2\over  z^4} +{dx_1^2+dx_2^2 \over z^2}+{dz^2\over
z^2} +{dy^2\over q^2}\right)    \br
&& e^{\phi_5}= g_5,~~~
 {\bar{C}}_{(3)}=  -{{L^3}\over  g_5 z^4}dt\wedge dx_1\wedge dx_2, ~~~
 {{B}}_{(2)}=  {{L^2}\over  qz^2}dt\wedge dy  \ ,
\eea
where 
${g}_5\equiv {g_a\over L^{5/2}}$ is 
 the  $5d$ string coupling constant. The $g_a$ is 10-dimensional coupling.
The corresponding $5d$ effective action can be presented as
\bea\label{eq3g1gg}
S_{5d}\sim\int\bigg[  e^{-2\phi_5}( \ast R +4(d\phi_5)^2   
-{ 1\over 2}(dB_{(2)})^2 +\ast{20\over L^2}) 
- {{\bar{m}}^2 \over 2}(B_{(2)})^2  
- { 1\over 2}(d{\bar{C}}_{(3)})^2  
-\ast{\bar{m}^2\over2}    
\bigg]  
\eea
where the parameters are related as $\bar{m}
=L^{5\over 2} m
\equiv {1\over g_5 L}$. The $B$ field is  explicitly massive. 
One may however rewrite the above action in terms of an 
axion field after the Hodge duality, $\ast dC_{(3)}= d C_{(0)}$ . 
The background \eqn{sol2vn3gg} is an exact  solution of 
the action \eqn{eq3g1gg} which has two cosmological constant terms
of opposite signs.

\subsection{The $4d$ Proca model}

After explicit
 compactification on $S^1\times S^5$ the 10-dimensional background  \eqn{sol2a}
reduces to the following 4-dimensional 
$Lif_4^{(2)}$ solution  
\bea\label{sol2vn3g}
&&ds^2=\bar{L}^2\left(- {dt^2\over  z^4} +{dx_1^2+dx_2^2 \over z^2}+{dz^2\over
z^2} \right)    \br
&& e^{\phi_4}=g_4, ~~~{\cal{A}}_{(1)}= - {{L}\over  z^2}dt  \ ,
\eea
where the  curvature radius $\bar{L}\equiv{L\over {g}_4}$ and
${g}_4\equiv {g_a\sqrt{q}\over L^3}$ is 
 the  $4d$ string coupling constant. 
The dilaton has got  constant background value. 
This background is a solution of the following Einstein-Proca effective
action, which 
directly follows from  $S^1\times S^5$
compactification of  the  Romans theory, 
\bea\label{eq3g1gp}
S_{4d}\sim \int \bigg[ \ast R -2 (d\phi_4)^2 
-{ e^{-2\phi_4}\over 2}(d{\cal{A}}_{(1)})^2 
- { m_4^2e^{2\phi_4}\over 2}({\cal{A}}_{(1)})^2 
+\ast{20e^{2\phi_4}\over L^2} 
  -\ast{5m_4^2 e^{4\phi_4}\over 2}    
\bigg]
\eea
where new mass parameter $m_4={2\over g_4 L}$.
The  action includes a Proca  field ({\it descending from the winding modes 
of the massive $B$-field in} \eqn{sol2a}). If we plug in the constant
value of the dilaton $g_4$, the action \eqn{eq3g1gp} further simplifies to
\bea\label{eq3g1g}
S_{4d}\sim \int \bigg[ \ast R 
-{ 1\over 2 g_4^2}(d{\cal{A}}_{(1)})^2 
- { 2\over L^2}({\cal{A}}_{(1)})^2 
+\ast {10 g_4^2\over L^2}   
\bigg].
\eea
We remind that the cosmological constant and Proca mass are very precisely
related in the action so as to admit Lifshitz $a=2$ vacua \eqn{sol2vn3g}. 
However the  action \eqn{eq3g1g}  cannot have  Schr\"odinger $a=3$ spacetime
  as  solutions.
Instead another Einstein-Proca action with a different cosmological constant, 
that follows from $S^5$ compactification of the fields in \eqn{sol2}, 
will allow $Sch_4^{(3)}$ solutions, 
see for the details \cite{Singh:2009tq,singh0866}. From this example
 we realise that, although $4d$ actions differ in only
the value of cosmological constants, but their
 $10d$ solutions have entirely different matter field contents.
These massive gauge-gravity  field models have been studied as 
effective  gravity models describing  holographic superconductor  
phenomena on  3-dimensional boundary  
\cite{bala,son,Hartnoll:2008kx}. We shall
provide  alternative $4d$ Lifshitz models which have two Maxwell potentials
and they are obtainable by exploiting massive T-duality
in the forthcoming sections.

\section{ $T^2$ dual of $(F1,D2,D8)$ system }
The 8-dimensional  type II supergravity 
 has a T-duality group $SL(2,R)\times SL(2,R)$ including the
massive version of supergravity \cite{hs2001}.
Under this massive duality symmetry the mass parameter $m$ 
(or dual 10-form $G_{(10)}$ flux) in 
Romans theory compactified on $T^2$ gets mapped into
$G_{(2)}$-flux (along 2-torus) in ordinary type-IIA compactified on $T^2$. 
Using this massive  duality we would like to map
 $Lif_4 \times R^1\times S^5$  vacua \eqn{sol2a} into the $G_{(2)}$-flux
vacua of ordinary type IIA theory. 

Let us choose $x_1$ and $x_2$ to be  the coordinates along  $T^2$. 
The   RR 2-form flux corresponds to having a nonvanishing 
constant  magnetic component  
\be
G_{flux}\equiv
{m }dx_1\wedge dx_2
={2 \over g_aL}dx_1\wedge dx_2
\ee
filling the entire $T^2$. 
(Note we must take mass  $m$ exactly equal to 
${2\over g_aL}$ as it is fixed earlier.) 
Using the duality map worked out in \cite{hs2001},   
we can now write down corresponding  vacua 
of ordinary type IIA theory,
\bea\label{sol2v}
&&ds^2=L^2\left( - {dt^2\over  z^4} +{z^2\over L^4 }(dx_1^2+dx_2^2)+{dz^2\over
z^2}  +{dy^2\over q^2} + d\Omega_5^2 \right) ,\br
&&e^\phi= {z^2\over L^2} g_a, ~~~
G_{(2)}= {4L^3\over z^5 g_a} dz\wedge dt + G_{flux}
,  \br 
&& B_{(2)}=  {L^2\over q z^2} dt\wedge dy \ ,
\eea 
Thus it could see that corresponding RR 1-form
gauge field has a most general form
\be
C_{(1)}=C_0 dt+C_1dx^1+C_2dx^2,~~~C_0=-{L^3\over g_a z^4},~~C_1= -
{1\over g_aL} x_2,~~C_2={1\over g_aL}  x_1
\ee 
which has both electric as well as magnetic components.
However no $C_{(3)}$ background is present, that
 is because after the duality all D2 branes (of massive-IIA) now 
reapper as  D0 branes in \eqn{sol2v}.
The   $B$ field   being  {\it massless}, as it is  in ordinary type IIA, 
 however  interacts (or gets nested) 
with nontrivial $G_{(2)}$ flux of D6-branes in the effective action. 
(Note all D8 branes  morphe into D6 branes after massive duality on $T^2$). 
The $B$ equation of motion  has a contributions from the
 interaction term in 4-form field strength
$G_{(4)}=dC_{(3)}+G_{(2)}\wedge B_{(2)}$ 
in type  IIA,
which for the above background \eqn{sol2v} contributes 
 a term like  $\sim  {1\over L^2} (B_{ty})^2$  in the action
due to nonvanishing $G_{(2)}$ flux. 
The equation of motion of   $C_{(3)}$ is trivially satisfied for $C_{(3)}=0$.
Thus $G_{(2)}$  has got both electric (D0) and magnetic (D6) components
all aligned along  noncompact directions in $5d$ spacetime in the above.
Thus the   Lifshitz background \eqn{sol2v} essentially
represents   $(F1, D0, D6)$ bound state in ordinary type IIA. 
{\it The D0 branes have reappeared back and F-strings are now massless.
This maybe called as un-higgsing phenomenon carried out by massive duality
on $T^2$. }

\section{Lower dimensional models}
\subsection{$4d$ Lifshitz theory and a quantum Hall system}
Let us now  consider the compact case where
$y\sim y +2\pi R_y$  in \eqn{sol2v}.  
The D6-branes will wrap around $S^1\times S^5$ completely 
and their  $G_2$ flux would
fill entire  $x_1-x_2$  plane. The massless F1-strings will be wrapping 
around  $y$ circle. The $4d$
 string coupling  goes to vanishing value in the UV (as $z\to 0$). 
Hence these vacua after compactification 
 give rise to  
$\widetilde{Lif}_4^{(2)}$ vacua (in string metric)

\bea\label{sol2vn}
&&ds^2=- {L^2dt^2\over  z^4} +{z^2\over L^2}(dx_1^2+dx_2^2)+{L^2 dz^2\over
z^2}     ,\br
&&e^{\phi_4}=  g_a {\sqrt{q}\over L^3}{z^2\over L^2}, ~~~
C_{(1)}= {1\over g_a}(-{L^3\over z^4}  dt - {x_2\over L} dx_1 + {x_1\over L} dx_2)
,  \br 
&& {\cal{A}}_{(1)}=  {-L^2\over q  z^2}dt  \ ,
\eea
with two distinct Maxwellian gauge fields ${\cal{A}}$ and $C$.
The scaling property of the  solution \eqn{sol2vn} under 
the $z\to\lambda z$ is
\bea\label{sca3}
&& t\to\lambda^2 t,~~ 
x_1\to\lambda^{-1} x_1,~~ 
x_2\to\lambda^{-1} x_2,~~ \br
&& g_{\m\n}\to g_{\m\n},~~~e^\phi\to \lambda^2 e^\phi, ~~~~
C\to\lambda^{-2} C ,~~~
 {\cal{A}} \to {\cal{A}} 
\eea
It  has a dynamical exponent of time as $a=2$, but crucially
has `negative' scaling exponent $(a_{x}=-1)$ for  spatial directions 
$x_1$ and $x_2$. It is quite plausible because D0 branes are delocalised 
over $x_1,~x_2$ plane as well as there is nontrivial
magnetic flux $G_{12}$. (The negative scaling exponent 
of spatial coordinates is usually associated with negative 
pressure along those direction 
in the CFT.)\footnote{The nonrelativisitic phenomenon  is expressed 
by the hamiltonian $H={(\vec{p})^2 \over 2m} + \cdots$, and  the scaling
property \eqn{sca3} implies that the quasi-particle masses 
must scale as $m\to \lambda^4 m$. Hence the particle
mass (number)  increases when we probe  bigger length scales 
(or at low energy). It should be
plausible for a quantum system exhibiting negative dynamical
exponent for spatial coordinates.}

The background \eqn{sol2vn} is  unique Lifshitz vacuum in the sense that
it describes two `electrically'  charged objects 
interacting with a magnetic flux. 
It can be checked  that the Lifshitz 
vacua \eqn{sol2vn} is indeed a solution of following massless 
$4d$ effective action 
\bea\label{eq3g}
S_{4d}\sim \int \bigg[ e^{-2\phi_4}\left( \ast R +4(d\phi_4)^2
-{q^2\over 2L^2}(d{\cal{A}}_{(1)})^2 +\ast {20\over L^2}\right)  
-{L^6\over 2q} G_{(2)}^2 - {qL^4\over 2} ({\cal{A}}_{(1)}\wedge G_{(2)})^2  
\bigg] \br 
\eea
where $\phi_4$ is 4-dimensional dilaton field and
field strength $G_{(2)}=dC_{(1)}$.  Note that 
the scaling of the  fields 
 in \eqn{sca3} gives rise to following  property 
of the action: 
\be
S_{4d}\to {S_{4d} \over \lambda^4} .
\ee
One thing to observe is that 
there is no mass term for gauge fields in the action \eqn{eq3g}. However
two Maxwell  fields in the action have Chern-Simons like interaction
between them. 
Hence  they give rise to two types of  charged objects with a Chern-Simons 
like interaction  $( {\cal{A}}\wedge dC)^2$ between them. 
The boundary Lifshitz  theory  lives over 2-dimensional spatial plane. 
By  constant  scaling of the  fields the action \eqn{eq3g} 
can be brought to a canonical form
\bea\label{eq3g1}
S_{4d}\sim \int \bigg[ \ast R -2(d\phi_4)^2
-{ e^{-2\phi_4}\over 2}(d{\cal{A}}_{(1)})^2 +\ast {20\over L^2}  e^{2\phi_4}  
-{1\over 2} (dC_{(1)})^2 - { e^{-2\phi_4}\over 2} 
({\cal{A}}_{(1)}\wedge dC_{(1)})^2  
\bigg] \br 
\eea
and corresponding $4d$ solution also can be written as 
\footnote{ The $4d$ string metric has a constant negative curvature. It does not include curvature singularity.}  
\bea\label{sol2vn3}
&&ds^2=L^2\left(- {dt^2\over  z^4} +{z^2\over L^4}(dx_1^2+dx_2^2)+{dz^2\over
z^2} \right)    ,\br
&&e^{\phi_4}=g_0{z^2\over L^2}, ~~~
C_{(1)}= {1\over g_0}(-{L^3\over z^4}  dt - {x_2\over L} dx_1 +{x_1\over L}dx_2),  \br 
&& {\cal{A}}_{(1)}=  -{L\over  z^2}dt  \ ,
\eea
where $g_0(\equiv g_a{\sqrt{q}\over L^3})$ is $4d$ coupling constant.
The parameter $q$  has altogether disappeared from  the 
solutions \eqn{sol2vn3} and it only determines the  coupling $g_0$. 
Since the dilaton runs towards strong coupling in the IR, this solution gives
valid CFT descrition only in the $z< L$ (near UV region). In the UV region the 
Lifshitz theory becomes almost free.

It is tempting to holographically 
relate the bulk  theory \eqn{eq3g1} to  some known nonrelativistic 
condensed matter phenomenon in a plane on the boundary,
 involving two distinct  (electrical) charges interacting
in presence of  constant magnetic field (or current), somewhat like in  quantum
Hall effect. There are no mass terms for  gauge fields hence 
the solution \eqn{sol2vn3} cannot
describe  superconducting or higgs phase. For the  bulk solution 
\eqn{sol2vn3}, 
the gauge interaction  terms in the action \eqn{eq3g1} is very crucial, but
it is of rather unusual type to motivate. However it is not uncommon
for such tensorial interactions to arise in bulk gravity (string) theory. 
The massless gauge-gravity
action  \eqn{eq3g1} certainly  represents a new critical phenomenon, 
but it is also related via `massive' T-duality to another action having 
 massive gauge fields  \eqn{eq3g1g} and describing
 superconducting phenomenon \cite{son, Hartnoll:2008kx}. 

Next the gauge field ${\cal{A}}$ specially  couples to the  dilaton 
hence it is  distinct as compared to the other gauge field $C$
 which is totally decoupled from
dilaton in  \eqn{eq3g1}. The former may thus have its 
origin in the Hall (excitaions) carriers.
As we understand from the quantum Hall effect that, 
there is a constant magnetic field uniformally spread over a  plane 
alongwith an electric field (EMF)  applied in one (say $x_1$) 
direction of the plane. 
Consequently a quantized Hall voltage (current) gets  
generated along another (here $x_2$) direction of the  planar system. 
From this analogy we understand that ${\cal{A}}_0$ can possibly
be the source of   Hall 
charges. In holography,  the boundary value of the time-component 
of bulk gauge field  represents 
charge sources in boundary. Furthermore,  ${\cal{A}}_0$ is
 essentially  a component of string $B$-field (wrapped on $y$ circle). 
Since $L^4\sim O(N)\gg 1$ and $L$  is an important overall parameter. We 
 find that the two gauge potentials  behave as
\be
{\cal A}_0\sim O(N^{1\over 4}), ~~~~ C_0\sim O(N^{3\over 4})\ee
Thus their respective carrier concentrations would  have $O(N^{1\over 2})$ 
difference. However
if we compare respective  Lorentz invariants in the action involving
 electric and magnetic fields, then
\bea
&&e^{-2\phi_4}(\partial_z{\cal A}_0)^2 \sim -{4g_0^2\over L^6}z^4 ,\br
&&(\partial_z{ C}_0)^2\sim -{16 g_0^2 \over L^6 } z^4 ,\br
&&(\partial_i{ C}_j)^2\sim {4g_0^2 \over L^6} z^4 \ .
\eea
It shows that, though the electric field contribution of 
$(\partial{\cal{A}})^2$ is comparatively weaker by a factor 
of $1/N$, but in the lagrangian \eqn{eq3g1} all gauge terms contribute
with equal strengths due to  varied dilatonic couplings. These
on-shell quantities do not diverge near the boundary.
Furthermore, in the bulk solution \eqn{sol2vn3} we can in fact choose a gauge
such that $C_0, C_1$ components are only nonvanishing alongwith ${\cal{A}}_0$. 
This implies existence of two  independent 
sources of charge (EMF) and an indepedent
electric current in the boundary theory.

\subsection{ Alternative  vacua with negative dynamical exponent}
It is obvious that the action \eqn{eq3g1} would  allow following 
Lifshitz vacua where dynamical exponent of time is instead  negative 
while two spatial coordinates scale positively. Let us
define a new holographic coordinate as
\be\label{def5}
u^2= {L^4\over z^2}
\ee
and by scaling  $t\to L^4 t$, we  obtain from  \eqn{sol2vn3} a new
kind of situation described by  
\bea\label{sol2vn3o}
&&ds^2=L^2\left(- { u^4 dt^2} +{1\over u^2}(dx_1^2+dx_2^2)+{du^2\over
u^2} \right)    ,\br
&&e^{\phi_4}=g_0{L^2 \over u^2}, ~~~~~
C_{(1)}= {1\over g_0}(-{u^4 \over L}  dt - {x_2\over L} dx_1 +{x_1\over L}dx_2),  \br 
&& {\cal{A}}_{(1)}=  -{u^2  L}dt  \ ,
\eea
which is valid in the $u> L$ (IR) region. This new  looking solution 
\eqn{sol2vn3o}  has negative
dynamical exponent for time ($a=-2$), such that
\bea\label{sca3o}
&& u\to\lambda u, ~~
t\to\lambda^{-2} t,~~ 
x_1\to\lambda x_1,~~ 
x_2\to\lambda x_2,~~ \br
&& g_{\m\n}\to g_{\m\n},~~~e^{\phi_4}\to \lambda^{-2} e^{\phi_4}, ~~~~
C\to\lambda^{2} C ,~~~
 {\cal{A}} \to {\cal{A}} .
\eea
Note  that the string
coupling tends to blow up at shorter length scales (UV), 
while at longer  (IR) scales it becomes almost a free theory. 
Thus the Lifshitz background \eqn{sol2vn3o} is suitable for 
describing a low energy Lifshitz theory
at large length scales $u>L$. (While at shorter  scales, $z<L$, the previous 
background \eqn{sol2vn3} is more suitable.)
In the present case the boundary field theory
may actually describe electrodynamics because coupling remains 
weaker at longer distances. 
\footnote{ The nonrelativisitic
phenomenon  is expressed 
by the hamiltonian $H={(\vec{p})^2 \over 2m} + \cdots$, and  the scaling
property \eqn{sca3o} implies that the quasi-particle masses 
must scale as $m\to m/\lambda^4 $. That is the quasi particle
mass (number) get reduced when  probed at large length scales 
(or with low energy) which appears to be
plausible in a quantum system exhibiting negative dynamical
exponent (for time) and having running coupling.}
 Once again there are two kinds of
(electric) charges interacting with a constant magnetic field such as in
quantum Hall systems near criticality.

\subsection{ $5d$ vortex model: $B\wedge G$ interaction}

A $5d$ effective action can be obtained by compactification 
of  type IIA theory on  the product spacetime  like
 ${\cal M}_5\times S^5$. We 
allow  fields to have dependence only on ${\cal M}_5$ coordinates,
such as we obtained in the  background  \eqn{sol2v}.
Upon consistent truncations,  and keeping only the  relevant field
content describing the equation \eqn{sol2v}, we  get to  following 
$5d$   action 
\bea\label{eq3}
S_{5}\sim \int \bigg[ e^{-2\phi_5}\left( \ast R +4(d\phi_5)^2
-{1\over 2}H_{(3)}^2 +\ast {20\over L^2}\right)  
-{1\over 2} G_{(2)}^2 - {1\over 2} (B_{(2)}\wedge G_{(2)})^2  
\bigg] \br 
\eea
The action has  second rank tensor with field strength
$H_{(3)}=dB_{(2)}$ and a vector field with 
$G_{(2)}=dC_{(1)}$, both interacting through a
 $B\wedge G$ like coupling. This interaction 
term will be important when  string-like
excitations (or flux tubes) 
couple with constant magnetic field in a transverse $2d$ plane
in  $4d$ CFT. As an example,
the equation of motion of the action \eqn{eq3}
are immediately solved by the following  solution
\bea\label{sol2v1l}
&&ds^2=- {L^2 dt^2\over  z^4}+{L^2 dy^2\over q^2 } +{z^2\over L^2}(dx_1^2+dx_2^2)+{L^2 dz^2\over
z^2} ,\br
&&e^{\phi_5}= g_0{z^2\over L^2} , ~~~
G_{(2)}= {4L^3\over g_0 z^5} dz\wedge dt + {1\over g_0L} dx_1\wedge dx_2 ,
\br&&~~~~ B_{(2)}=  -{L^2\over q z^2}dt\wedge dy, 
\eea 
where $g_0=g_a/ \sqrt{L^5}$ is $5d$ coupling constant and $q$ is 
arbitrary  having dimensions of length. The $5d$ spacetime 
has constant negative curvature. 
The  coupling remains weak
in the UV region. The solution represents an uniform string like
excitation  extended in 
$y$ direction and a constant magnetic $G$-flux $\sim  {1\over g_0L}$
 in the transverse $x_1-x_2$ plane.

The above solution \eqn{sol2v1l} is however  well behaved
only in the UV region where $z<L$.
But, by using the transformation \eqn{def5} one can transform it
into the IR region ($u>L$)
\bea\label{sol2v2l}
&&ds^2=L^2\left(- {dt^2\over  u^{-4}}
+{dx_1^2+dx_2^2 \over u^2}+{ du^2\over u^2} +{ dy^2\over q^2 } 
\right)\br
&&e^{\phi_5}= g_0{L^2\over u^2} , ~~~
G_{(2)}= {4u^3\over g_0 L} du\wedge dt + {1\over g_0L} dx_1\wedge dx_2 ,
\br&&~~~~ B_{(2)}=  -{L^2u^2\over q }dt\wedge dy, 
\eea 
where coupling remains weaker in the deep IR region. 
The dynamical exponent of time is given by $a=-2$. The charge
sources are also  present because the time-component $C_0$
is nontrivial. There is a constant magnetic background too, but
there are also extended string  (vortex) like objects present which couple to
 $B_{0y}$. It would be worthwhile to do a detailed analysis 
of the  nonrelativistic
CFT on the boundary. We hope to come back to this topic
in a future investigation.

\section{ (W, M2) brane system}
The (F1,D2,D8) system \eqn{sol2a} cannot be uplifted to M-theory at strong 
coupling as it is a solution of Romans theory for which we do not have
11-dimensional interpretation. Also because 
$L\sim {1\over m g_a l_s}\to 0$ at strong coupling
the spacetime becomes highly curved. On the other hand,  
we  obtained (F1,D0,D6) in eq.\eqn{sol2v} as the solution of
 ordinary type IIA supergravity. For the massless type IIA theory 
there exists an M-theory description \cite{Witten:1995ex}.
For the  vacua \eqn{sol2v} the string coupling becomes strong in 
the IR region $(z\to\infty$)
which usually is the case for CFT$_3$ (in relativistic cases also)
 \cite{itzhaki}.   
At strong coupling a natural explanation of the theory 
should be found by studying these
solutions  in M-theory (with $R_{11}=g_a^{2/3}l_p,~l_p=g_a^{-1/3}l_s$) 
\be\label{m11}
ds^2_{11d}=e^{4\phi\over 3}(dx_{11}+A)^2 +
e^{-2\phi\over 3}ds_{10d}^2 \ee
where $x_{11}\sim x_{11} +2\pi R_{11}$ is the eleventh coordinate.
Correspondingly $11$-dimensional supergravity    vacua can be described as
(setting $L=1$ for simplicity)
\bea\label{sol2v1}
&&ds^2_{11d}={1\over  z^{4/3}}\left(
- {dt^2\over  z^4}  +z^2(dx_1^2+dx_2^2)
+{dz^2\over z^2} 
+{z^4}(dx_{11}+\omega)^2+{1\over w^2} d\psi^2 + d\Omega_5^2  \right),\br
&& ~~~~ C_{(3)}=  {1\over w z^2}dt\wedge dx_{11}\wedge d\psi  \ ,
\eea 
where fiber 1-form is 
$$ \omega_{(1)}=-z^{-4} dt  -x_2dx_1 + x_1dx_2 .$$
The background in \eqn{m11} represents a conformal 
$Lif_5^{(2)} \times S^1_w\times S^5$ vacua in M-theory  which
includes  circle $x_{11}$  fibered over a $Lif_4$ base geometry:  
$- {dt^2\over  z^4}  +z^2(dx_1^2+dx_2^2)
+{dz^2\over z^2}$. The fiber 1-form is special in that it has
both time as well as  magnetic  components.
The magnetic components can be thought off as
 an uplift of  D6-branes in \eqn{sol2v}. (Note 
D6 branes will have only geometric interpretation when lifted to M-theory
 where magnetic type IIA 1-form
becomes the fiber along M-theory circle.). 
The  M2-branes   wrap $(x_{11},\psi)$ completely.
There is  large 
anisotropy along $x_1,~x_2,~x_{11}$ directions of the  Lifshitz spacetime; 
\bea\label{sca4}
&& t\to\lambda^2 t,~~ 
x_1\to\lambda^{-1} x_1,~~ 
x_2\to\lambda^{-1} x_2,~~ 
x_{11}\to\lambda^{-2} x_{11},~~ \br
\eea
under which the metric conformally scales  as 
$g_{\m\n} \to\lambda^{-4/3} g_{\m\n}$ while 
the 3-form $C \to\lambda^{-2} C$.

\section{Summary}

We have shown  
that for the  Romans' type IIA  supergravity  the $(F1,D2,D8)$  brane
 configuration  gives rise to the Lifshitz vacua   
$Lif_4^{(2)}\times {R}^1\times S^5$. For this Lifshitz 
solution  the fundamental  strings have to be  massive with
$m= {1\over g_a L l_s } $, and such that $L\gg 1$ and $g_a \ll 1$.
A compactification of the massive type IIA theory 
on $S^1$ relates  $9$-dimensional  $Lif_4^{(2)}\times S^5$  
background with  corresponding Scherk-Schwarz reduced
type IIB Lifshitz vacua compactified on a dual circle. This duality map is
known as  `massive' T-duality \cite{berg}. It has been known in litrature that
 $Lif_4^{(2)}\times S^1\times S^5$ spacetime 
  is  constituted by axion-flux  $(W,D3,D7)$ system
 in type IIB string theory. 
The presence of D7-branes  primarily requires  constant axion flux
switched on along $S^1$ in these solutions. 
None of these solutions preserve any supersymmetry.

Further,  using  `massive' T-duality symmetry in eight dimensions, 
we have  related the  Lifshitz vacua in Romans theory 
 to yet another  $\widetilde{Lif}_4^{(2)}\times R^1\times S^5$
 vacua which is constituted by  $(F1,D0,D6)$  brane system
of ordinary  type IIA theory. But these latter vacua have  negative 
dynamical exponents along  two  CFT directions. 
The interesting observation is that this $(F1,D0,D6)$ 
solution upon exlicit compactification along $S^1\times S^5$ gives rise to 
$4d$ Lifshitz vacua with two distinct types of electric  charges
supported by a constant magnetic flux (or current), all entirely 
living over $2d$ plane at the boundary. We speculate 
that this may  represent phenomena akin to  quantum Hall systems. 
We also have  obtained  M-theory uplift of the $a=2$
Lifshitz solutions.

\vskip1cm
\leftline{\bf Acknowledgments:} I am thankful to  
the ICTP, Trieste for the associateship support and 
the ASC-LMU, Munich  for kind  hospitality where parts of this work 
are carried out.


\end{document}